\documentclass[aps,prl,twocolumn,groupedaddress]{revtex4}
\pdfoutput=1
\usepackage{graphicx}
\usepackage{amsmath,amsfonts,amssymb}

\usepackage{color}
\begin{document}
\title{From DeWitt initial condition to
Cosmological Quantum Entanglement}

\author{Aharon Davidson} 
\email{davidson@bgu.ac.il}
\homepage[Homepage: ]{http://www.bgu.ac.il/~davidson}
\affiliation{Physics Department, Ben-Gurion University
of the Negev, Beer-Sheva 84105, Israel}
\author{Tomer Ygael}
\email{tomeryg@post.bgu.ac.il}
\affiliation{Physics Department, Ben-Gurion University
of the Negev, Beer-Sheva 84105, Israel}

\date{April 04, 2015}

\begin{abstract}
At the classical level, a linear dilaton action offers an
eternal inflation evolution governed by the unified (cosmological
constant plus radiation) equation of state $\rho-3P=4\Lambda$.
At the quantum mechanical mini superspace level, a 'two-particle'
variant of the no-boundary proposal, notably 'one-particle' energy
dependent, is encountered.
While a (gravity-anti-gravity) GaG-odd wave function can only
host a weak Big Bang boundary condition, albeit for any $k$, a
strong Big Bang boundary condition requires a GaG-even entangled
wave function, and singles out $k=0$ flat space.
The locally maximal values for the cosmological scale factor
and the dilaton form a grid
$\{a^2,a\phi\}\sim\sqrt{4n_1+1}\pm\sqrt{4n_2+1}$.
\end{abstract}

\maketitle

\noindent \textbf{Introduction}

The idea that the Newton constant may have changed
its sign during the very early universe is not new.
Originally, it was presented \cite{Linde} as a possible
consequence of the interaction of geometry with
electro/nuclear matter.
With the early universe in mind, it has been recently suggested
\cite{Bars} that anti-gravity may have played a crucial role
in taming the classical Big Bang singularity by curing the
geodesic incompleteness problem.
A central ingredient in this theory is the underlying conformal
Weyl symmetry (see \cite{Jackiw} for rermarks).
The fact is, however, that anti-gravity can be regarded
a natural companion of gravity in almost any scalar-tensor
theory \cite{ScalarTensor}.
Such an intimate relation can even be elevated to the level
of a gravity $\leftrightarrow$ anti-gravity (GaG) discrete
symmetry \cite{Quiros}.
The latter originally includes a metric change of sign,
which is hereby traded for a cosmological scale factor
change of sign, causing our variant GaG symmetry to make
its appearance only at the level of the mini superspace (and
the classical equations of motion).

Starting from a linear dilaton gravity Lagrangian, we first
establish the classical mirror cosmology, and demonstrate
how spatially symmetric mirror-gravity deflation can smoothly
evolve into eternal gravity inflation.
The accompanying singular Big Bang (BB) solution decouples
gravity from anti-gravity.
Reflecting the fact that the Ricci scalar serves as a constant
of motion, the emerging classical equations of state signal the
presence of a mandatory (not put ad-hoc) radiation companion
to the cosmological constant.
In the quantum mechanical picture, at least within the
framework of the mini superspace model, we encounter a
'two-particle' variant of the Hartle-Hawking
no-boundary proposal \cite{HH}.
While the Hamiltonian constraint is fully respected, the
scheme still allows for a 'one-particle' energy dependence,
highly resembling Vilenkin's tunnelling wave function
approach \cite{Vilenkin} with radiation included. 

Quantum mechanically, the Big Bang can never stay out of
the game, but here, within the framework of a singularity-free
boundary proposal, it is accompanied by a non-trivial
interplay.
While a GaG-odd/BB-even entangled wave function can only host a weak
DeWitt boundary condition \cite{DeWitt} at the Big Bang, albeit for
any spacial curvature $k$, a strong DeWitt boundary condition
requires a GaG-even/BB-odd entangled wave function, and furthermore
singles out the $k=0$ flat space option.
The classically problematic (under anisotropic fluctuations
\cite{Catastrophe}) GaG transition does not seem to leave
pathological imprints on the outgoing Wheeler-DeWitt (WDW)
wave function.
The highlight of our analysis is the emergence of a grid structure
for the locally most probable cosmological scale factor
and the inverse Newton constant.

\medskip\noindent\textbf{Dilaton cosmology preliminaries}

In its simplest form, devoid of a Brans-Dicke \cite{BD} kinetic term
($\omega_{BD}=0$),
dilation gravity is formulated by means of the action principle
\begin{equation}
		{\cal I}=-\int \left( \phi {\cal R}
		+V(\phi)\right)\sqrt{-g}~d^4 x ~.
\end{equation}
The gravity ($\phi>0$) anti-gravity ($\phi<0$) transition,
depending on the potential, is not necessarily singular.
The GR limit, when exists \cite{GRlimit}, is associated with
$\phi(x)$ developing a positive vacuum expectation value
$(8\pi G)^{-1}$.
The cosmological field equations which govern the
Friedmann-Lema"tre-Robertson-Walker (FLRW) metric
\begin{equation}
	ds^2=-dt^2+a^2(t)
	\left(\frac{dr^2}{1-kr^2}+r^2d\Omega^2\right)
\end{equation}
can be compactly arranged into
\begin{subequations}
\begin{eqnarray}
	&&\frac{\dot{a}}{a}\dot{\phi}
	+\frac{\dot{a}^2+k}{a^2}\phi
	=\frac{1}{6}V(\phi) ~,
	\label{eqa}\\
	&& \frac{\ddot{a}}{a}
	+\frac{\dot{a}^2+k}{a^2}
	=\frac{1}{6}V^{\prime}(\phi) ~.
	\label{eqb}
\end{eqnarray}
\end{subequations}
A deeper physical insight can be gained by reconstructing
the Klein Gordon equation
\begin{subequations}
\begin{eqnarray}
	&\displaystyle{\ddot{\phi}+3\frac{\dot{a}}{a}\dot{\phi}
	+V_{eff}^{\prime}(\phi)=0 }~,&
	\label{KG}\\
	& \displaystyle{V_{eff}(\phi)=\frac{1}{3}\int \left(
	\phi V^{\prime}(\phi)-2V(\phi)
	\right) d\phi} ~.&
\end{eqnarray}
\end{subequations}
$V_{eff}(\phi)$ is not sensitive to
$V(\phi)\rightarrow V(\phi)+\lambda \phi^2$.
The inclusion of a Brans-Dicke kinetic term would only modify its scale
$1/3\rightarrow 1/(3+2\omega_{BD})$.

\medskip
\noindent\textbf{Inflation/radiation interplay}

Let our starting point be a particular dilaton gravity theory
characterized by a linear scalar potential, namely
\begin{equation}
	V(\phi)=4\Lambda \phi ~,
	\quad \Lambda>0 ~.
	\label{linearV}
\end{equation}
Note that such a dilaton theory does not have an analogous
$f({\cal R})$ gravity theory \cite{f(R)} because the dictionary
usually used to extract
$\phi({\cal R})$, namely ${\cal R}+V^{\prime}(\phi)=0$,
turns a constraint in this case.
Associated with the linear potential eq.(\ref{linearV}) is
the quadratic 'wrong sign' (no ground state) effective potential
\begin{equation}
	V_{eff}(\phi)=-\frac{2}{3}\Lambda \phi^2 ~,
	\label{Veff}
\end{equation}
which governs the Klein-Gordon evolution of the dilaton
field.
It constitutes a vital ingredient for a realistic spontaneously
induced GR theory (for comparison, Starobinsky's
${\cal R}+{\cal R}^2$ type gravity \cite{Starobinsky} comes
with $V_{eff}\sim+(\phi-v)^2$).

The scale factor evolution can be directly derived from
the now $\phi$-independent non-linear differential
equation eq.(\ref{eqb}).
The bouncing solution
\begin{subequations}
\begin{eqnarray}
	&&\displaystyle{a^2 (t)-\frac{3k}{2\Lambda}=
	A^2 \cosh \omega t~,}
	\label{a}\\
	&& \displaystyle{a(t)\phi(t)=B^2 \sinh \omega t~,
	\quad  \omega ^2=\frac{4}{3}\Lambda ~,}
	\label{phi}
\end{eqnarray}
\end{subequations}
attracts particular attention because of the special case
$A^2=\frac{3k}{2\Lambda}$ which has been used by
Hartle-Hawking in the no-boundary proposal \cite{HH}
formulation (see \cite{bounce} for stringy and loop
bouncing solutions).
The other solution, characterized by a Big Bang singularity,
is obtained by switching the hyperbolic trigonometric
functions.
While Ricci scalar serves as a constant of motion
\begin{equation}
	{\cal R}=-4\Lambda ~,
\end{equation}
all other curvature scalars are non-singular solely for the
bouncing solution provided
$a_0^2=\frac{3k}{2\Lambda}+A^2>0$.

It is convenient to recast the gravitational Jordan frame field
equations into the standard general relativistic format
${\cal R}_{\mu\nu}-\frac{1}{2}g_{\mu\nu}{\cal R}=
{\cal T}_{\mu\nu}$, squeezing all floating around pieces
into an effective energy/momentum tensor on the rhs.
The corresponding FLRW cosmology would then involve
an effective energy density $\rho=3(H^2+k/a^2)$
accompanied by an effective pressure $P$.
Together, these two close upon  the energy/momentum
conservation law $\dot{\rho}+3H(\rho+P)=0$.
A closer inspection reveals that the entire cosmic evolution,
from the Big Bang to the de-Sitter limit, is governed by the
unified equation of state
\begin{equation}
	\rho-3P=4\Lambda ~,
	\label{state}
\end{equation}
being a direct consequence of the linear potential dilaton
field equation.
The differential version of the equation of state, namely
$d\rho-3dP=0$, signals the presence of an evolving
radiation component.
To be specific, associated with the scale factor eq.(\ref{a})
is the total energy density
\begin{equation}
	\rho(t)=\Lambda
	\left(1+\frac{\frac{9k^2}{4\Lambda^2}
	-A^4}{ a^4(t)}\right)~.
	\label{rho}
\end{equation}

As long as the underlying Lagrangian is GaG-odd, that is
$V(-\phi)=-V(\phi)$, the corresponding effective potential
is by construction GaG-even $V_{eff}(-\phi)=V_{eff}(\phi)$.
In turn, the equations of motion stay intact under
$\phi\rightarrow -\phi$.
Such a symmetry, unaffected by the possible presence
of a Brans Dicke kinetic term, suggests itself as the tool
for synchronizing the Big Bang creation with the
gravity anti-gravity transition.
With this in mind, \emph{mirror gravity} is perhaps
a better name for the GaG anti-symmetric gravity
companion.
After all, an optical/electromagnetic and even a
gravitational \cite{Clifton} mirror image is a reflected
duplication of an object that appears identical but reversed.

\medskip\noindent\textbf{Bifurcated mini superspace}
 
 Mirror gravity admits an elegant diagonalization at the
 mini superspace level, where the GaG odd linear potential
 eq.(\ref{linearV})  leaves a distinctive fingerprint.
Following the standard procedure of integrating out over
the maximally symmetric space, that is
 \begin{equation}
	\int{\cal L}\sqrt{-g}~dtd^{3}x~ \longrightarrow ~
	\int {\cal L}_{mini}~dt~,
\end{equation}
and up to a total derivative, the mini superspace Lagrangian is
given by
\begin{equation}
	{\cal L}_{mini}=-\frac{6}{n}
	\left(a\phi\dot{a}^2+a^2 \dot{a}\dot{\phi}\right)
	+n\phi(6k a-4a^3\Lambda) ~.
	\label{Lmini}
\end{equation}
Note that the lapse function $n(t)$ has been revived to keep
track of the underlying diffeomorphism.
By supplementing $\phi\rightarrow -\phi$ by $a\rightarrow-a$
the GaG symmetry has made its appearance at the mini
superspace level.
A new pair of GaG parity conjugate variables, namely
\begin{equation}
	x_{\pm}=\frac{a}{2}(a \pm \phi) ~,
\end{equation}
is then invoked to diagonalize the quadratic kinetic term
($\phi$ is dimensionless in our $8\pi G=1$ notations).
Consequently, the above GaG-odd mini superspace Lagrangian,
with $\phi\rightarrow -\phi$ (or $a\rightarrow -a$) reading
now $x_+\leftrightarrow x_-$, gets decomposed into
${\cal L}={\cal L}_{+}-{\cal L}_{-}$.
Note that such a decomposition is not possible in the presence
of a kinetic Brans-Dicke term $\omega_{BD}\neq 0$.
Using the more convenient Hamiltonian language,
we face
\begin{equation}
	{\cal H}_{mini}=n\left({\cal H}(x_{+},p_{+})
	-{\cal H}(x_{-},p_{-})\right)~.
\end{equation}
The upside-down harmonic oscillator building block 
 \begin{equation}
	{\cal H}(x,p)=\frac{p^2}{12}
	+U(x) ~,~~U(x)=6kx-4\Lambda x^2 ~,
	\label{Hmini}
\end{equation}
resembles in a way the Hartle-Hawking Hamiltonian
\begin{equation}
	{\cal H}_{HH}(a,p)\propto \frac{p^2}{4~}
	+k a^2-\frac{1}{3}\Lambda a^4 ~.
	\label{HH}
\end{equation}
The two almost decoupled 'one-particle' physical systems
communicate with each other solely by means of the overall
Hamiltonian constraint $ {\cal H}_{mini}=0$.
Their associated conserved energies, $E_{+}$ and $E_{-}$
respectively, must therefore obey $E_{+}-E_{-}=0$.
Thus, in contrast with the Hartle-Hawking prescription, a bit
closer to Vilenkin approach, a non-vanishing energy parameter
$E_{+}=E_{-}=E$ makes it appearance via
\begin{equation}
	{\cal H}(x_{+},p_{+})
	={\cal H}(x_{-},p_{-})=E ~.
\end{equation}
The corresponding mechanical analogue involves two identical
non-interacting point particles, carrying the one and the same 
energy $E$, whose one dimensional motion is governed by the
$U(x)$ potential.
\begin{figure}[h]
	\includegraphics[scale=0.4]{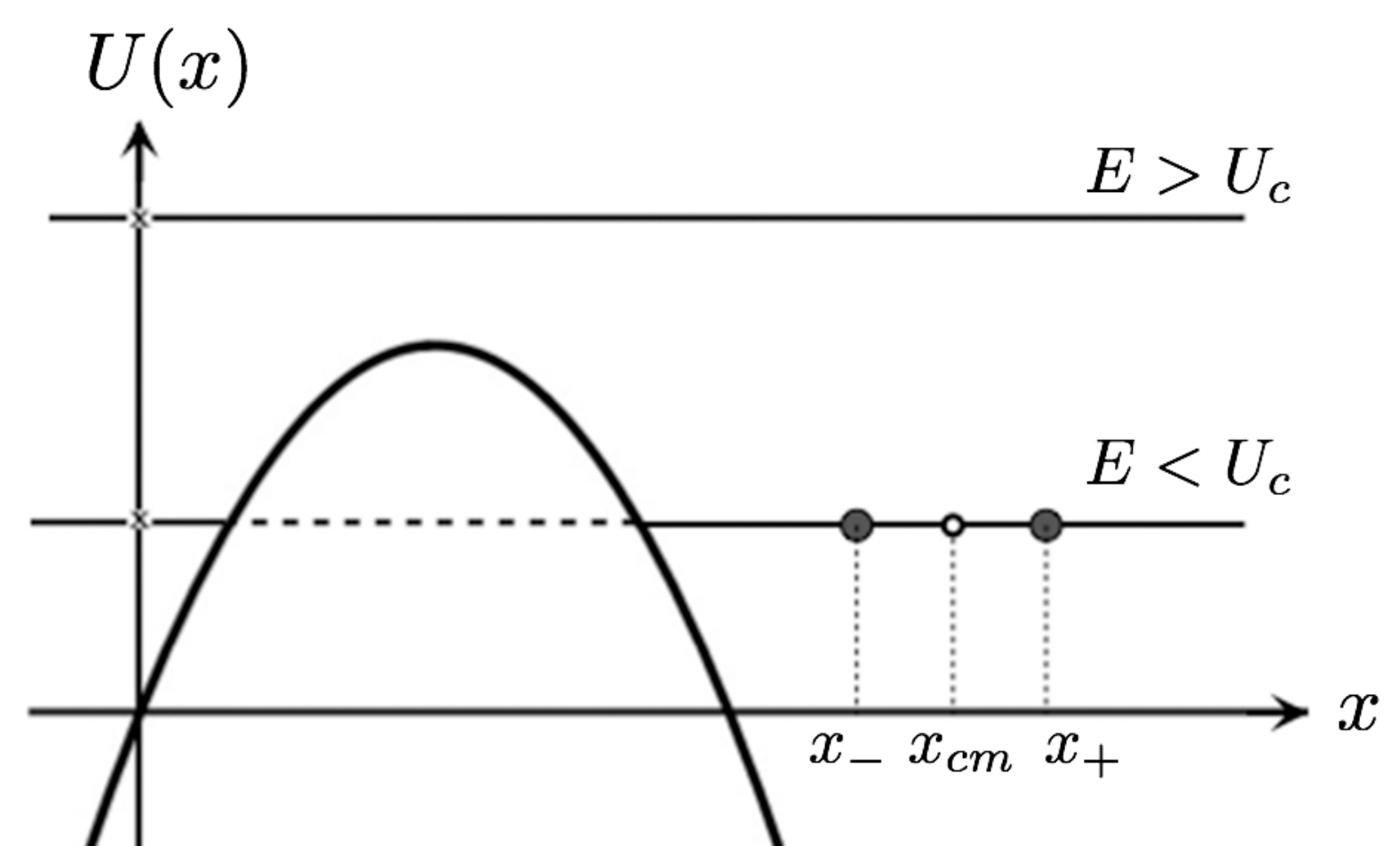}
	\caption{The motion of two identical non-interacting
	point particles, carrying a total energy $E$ each,
	is governed by the one-particle potential $U(x)$.
	While their center of mass location represents the
	cosmological scale factor squared $a^2 (t)$, their
	crossing point marks the gravity anti-gravity transition.}
	\label{U.pdf}
\end{figure}

The most general classical solution of the associated field equations
is given by
 \begin{equation}
	\left[\begin{array}{c}x_+(t) \\x_-(t)\end{array}\right]=
	\frac{9k^2}{4\Lambda}+
	\left[\begin{array}{cc}c^+_{+} & c^-_{+} 
	\\c^+_{-}  & c^-_{-} \end{array}\right]
	\left[\begin{array}{c}e^{+\omega t}\\
	e^{-\omega t}\end{array}\right] ~.
\end{equation}
While fixing the 'one-particle' energy
\begin{equation}
	E=\frac{9k^2}{4\Lambda}-16\Lambda c^+_{+}c^-_{+}
	=\frac{9k^2}{4\Lambda}-16\Lambda c^+_{-}c^-_{-}
\end{equation}
counts for two relations among the various $c$-parameters,
fixing the origin of time is translated into scaling
$c^+_{\pm}$ and inversely scaling $c^-_{\pm}$, thus leaving
the $c^+_{+}/c^+_{-}$ ratio as the second physical parameter.

The bouncing solution specified by eqs.(\ref{a},\ref{phi})
is associated with an energy level
\begin{equation}
	E=\Lambda\left(
	\frac{9k^2}{4\Lambda^2}-A^4\right)
	+\Lambda B^4~.
	\label{E}
\end{equation}
The first term is immediately recognized as the
strength of the radiation energy density eq.(\ref{rho}).
A positive radiation energy density guaranties $E\geq 0$.
If furthermore $A^2>B^2$, both point particles, each
carrying energy $E<U_{c}=\frac{9k^2}{4\Lambda}$, move
from right to left towards the potential hill.
As they hit the turning point, the first at some time $t_1$
and the other a bit later at $t_2$, they must cross each other
in between.
Such a crossover, with $x_{+}-x_{-}$ changing sign, signals
the GaG transition.
If on the other hand $B^2>A^2$, the two point
particles, each carrying now a larger energy $E>U_{c}$,
move in opposite directions above the potential hill.
Like before, their unavoidable crossing point fixes the minimal
$a_0^2$ at the GaG transition.
In both cases, the center of mass location stays positive definite,
causing no classical Big Bang singularity.

\begin{figure}[h]
	\includegraphics[scale=0.32]{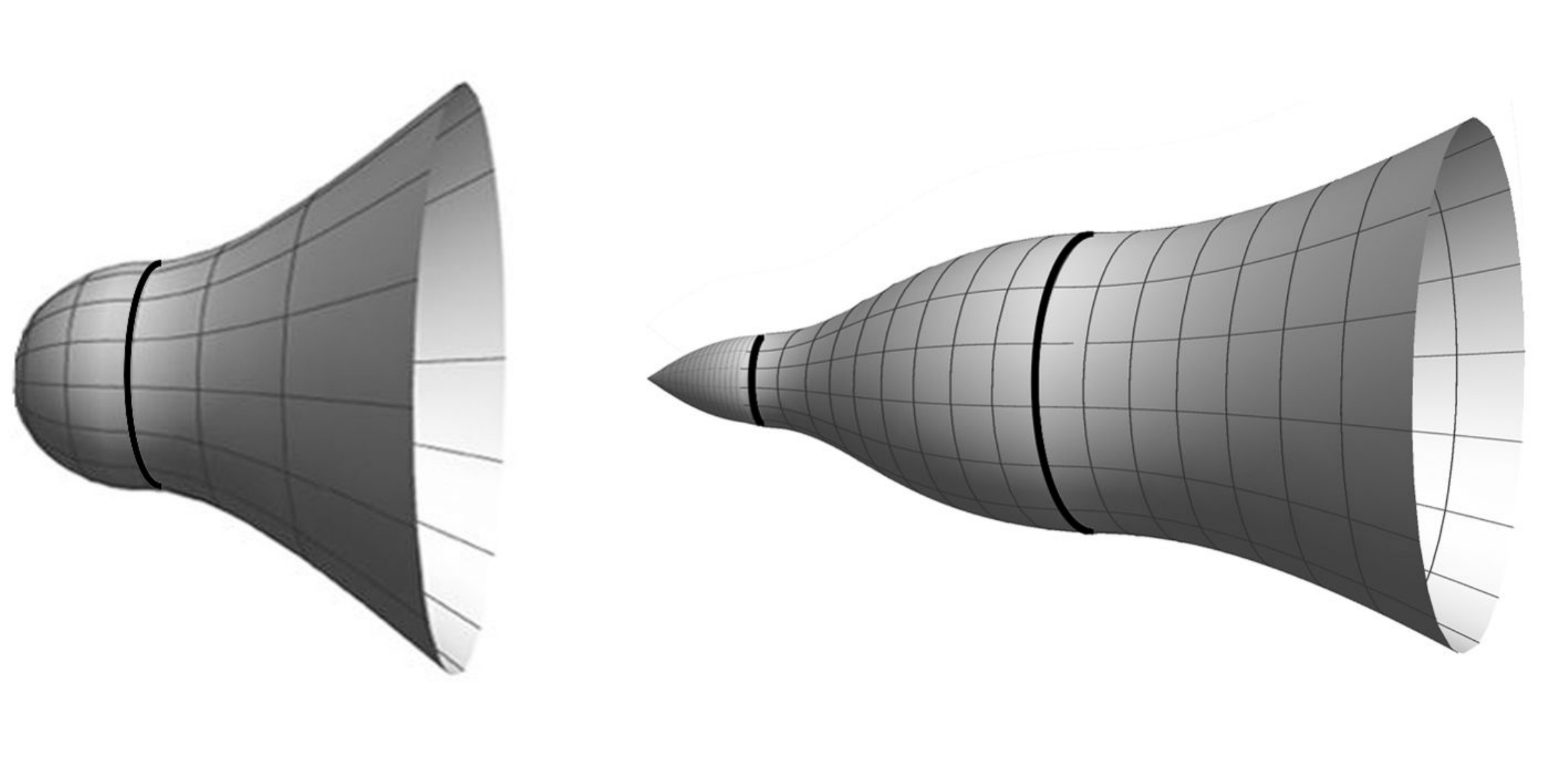}
	\caption{$E$-dependent variant (right) of the
	no-boundary proposal (left).
	The intermediate Euclidean part ceases
	to exist for $E\geq U_c$.
	The tip singularity is diffused by weak/strong
	DeWitt boundary condition.}
	\label{NoBoundary}
\end{figure}

The fact that the energy formula  is only sensitive to
$A^4$, rather than to $A^2$, implies that the two classically
disconnected solutions $a_{\pm}^2 (t)= \frac{3k}{2\Lambda}
\pm A^2 \cosh \omega t$ share the one and the same energy
$0<E<U_{c}$.
The $a_{-}(t)$ branch describes a Big Bang created baby universe
that does not have a chance to grow forever.
Quantum mechanically, however, the story is by far more interesting
as the Euclidean sector $a_{E}(\tau)$ connects the
embryonic and the expanding Lorentzian regions to eventually
constitute the piecewise glued sharp edge manifold depicted in
fig.(\ref{NoBoundary}).
The existence of an embryonic stage in the quantum history
was originally introduced by Vilenkin who supplemented the
Hartle Hawking proposal by an ad-hoc radiation energy density.
At any rate, the quantum mechanical picture is by far richer.
The $E$ degree of freedom allows to (i) Go beyond the Euclidean
regime, (ii) Let $k\leq 0$ into the game, and most importantly
(iii) Consistently generalize the DeWitt boundary condition.

\noindent\textbf{Quantum entanglement and a quantum grid}

The Hamiltonian constraint destroys the explicit time dependence
of the universe wave function.
For a time independent wave eigenfunction, of the separate
variables form $\Psi_E (x_{+},x_{-})=\psi_E (x_{+}) \psi_E (x_{-})$,
the corresponding WDW equation splits into
\begin{equation}
	\left[-\frac{\hbar^2}{12}
	\frac{\partial^2}{\partial x^2}
	+U(x)-E \right] \psi_E (x)=0
	\quad\text{for} ~x=x_{\pm}~.
	\label{WDW}
\end{equation}
Each individual differential equation is not a
legitimate Schrodinger equation by its own rights since the
(absence of) time evolution is controlled by the full system.

\begin{figure}[h]
	\includegraphics[scale=0.43]{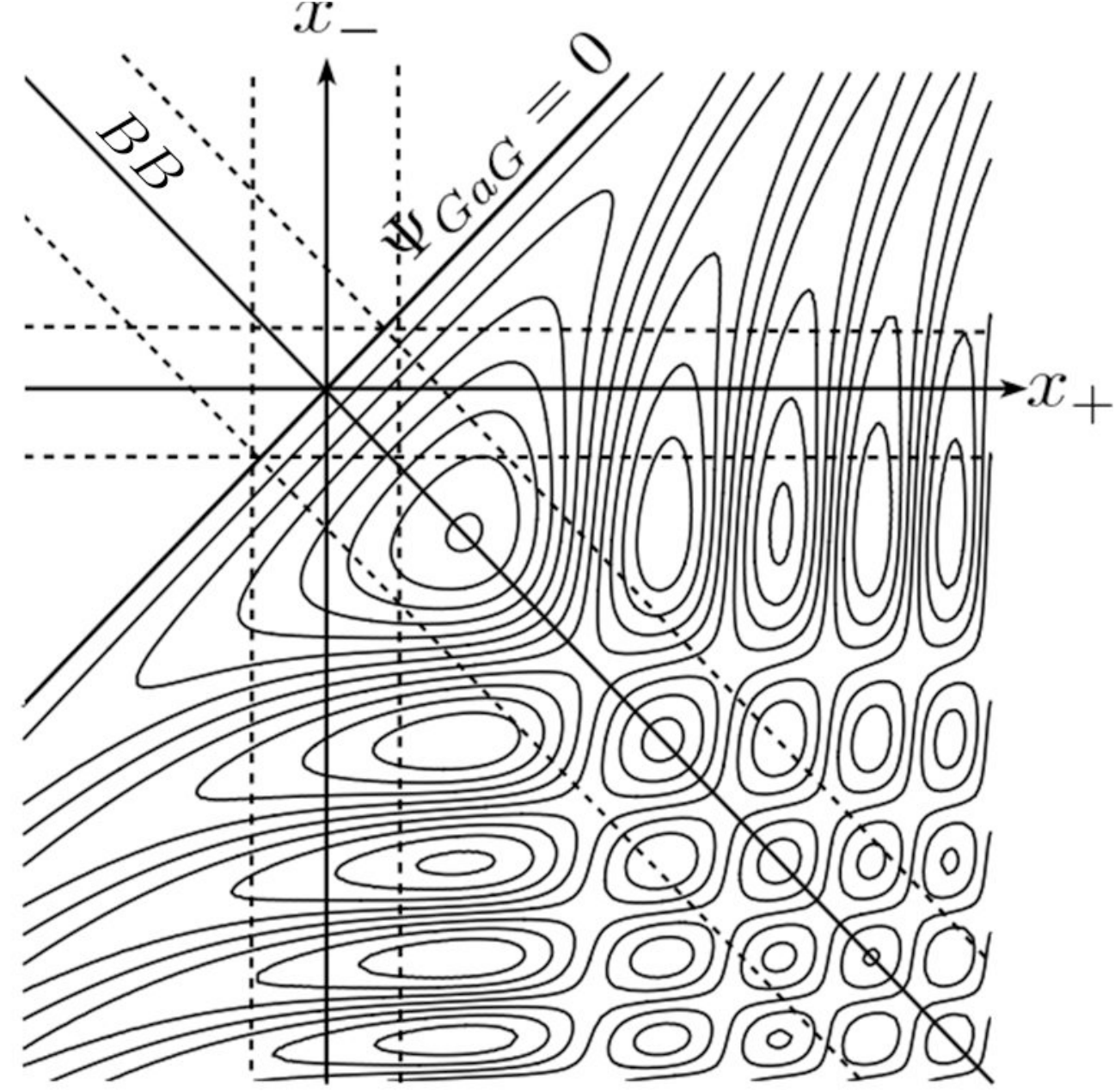}
	\caption{$|\Psi_W|^2$ contour plot
	(solid axes $k=0$, dashed axes $k=\pm 1$):
	$\Psi_W$ is GaG-odd/BB-even, so the anti-gravity
	sector can be dismissed.
	Notice the grid structure at the 4th quarter.}
	\label{ContourGaG}
\end{figure}

The one-particle Schrodinger equation admits two independent
solutions.
They are
\begin{eqnarray}
	&&R(x)=D_{-\frac{1}{2}(1-is)}
	\left[\frac{2}{\sqrt{\hbar}}(i-1){(3\Lambda)}^{\frac{1}{4}}
	(x-{ \frac{3k}{4\Lambda}})\right]~~~\\
	&&L(x)=D_{-\frac{1}{2}(1+is)}
	\left[\frac{2}{\sqrt{\hbar}}(i+1){(3\Lambda)}^{\frac{1}{4}}
	(x-{ \frac{3k}{4\Lambda}})\right]~~~
\end{eqnarray}
$D_n (x)$ stands for the Weber parabolic cylinder function,
and $s(E)=\frac{1}{\hbar}
\sqrt{\frac{3}{\Lambda}}\left(E-\frac{9k^2}{4\Lambda}\right)$.
On simplicity grounds we stick to fixed $E$, but keep in
mind wave packet solutions \cite{Kiefer} as well.
Regarding boundary conditions, primarily invoked to relax the
classical Big Bang singularity, two options arise:

\smallskip
\noindent \emph{(i) Weak DeWitt initial condition
$\psi(0,0)=0$:}
As the cosmological scale factor approaches zero,
$\psi\rightarrow 0$  for any finite value of $\phi$.
Such a requirement can be naturally embedded, notably
for any value of $k$, and in accord with the global symmetries
of the mini superspace Hamiltonian, within the strong GaG boundary
condition $\psi(x,x)=0$.
Up to a normalization factor, the entangled WDW wave function
takes the real by construction form
\begin{equation}
	\Psi_W (x_{+},x_{-})= \frac{1}{\sqrt{2}}
	\left(R(x_+)L(x_-)-R(x_-)L(x_+)\right)
\end{equation}
Reflecting the GaG-odd nature of $\Psi_W$, the anti-gravity sector
can be eliminated.
Roughly speaking, the classical analog is the
$\{a^2(t)\sim\sinh\omega t,~a(t)\phi(t)\sim \cosh\omega t\}$-type
solution.
The $|\Psi_W|^2$ contour plot is depicted in Fig.(\ref{ContourGaG}).

\begin{figure}[h]
	\includegraphics[scale=0.43]{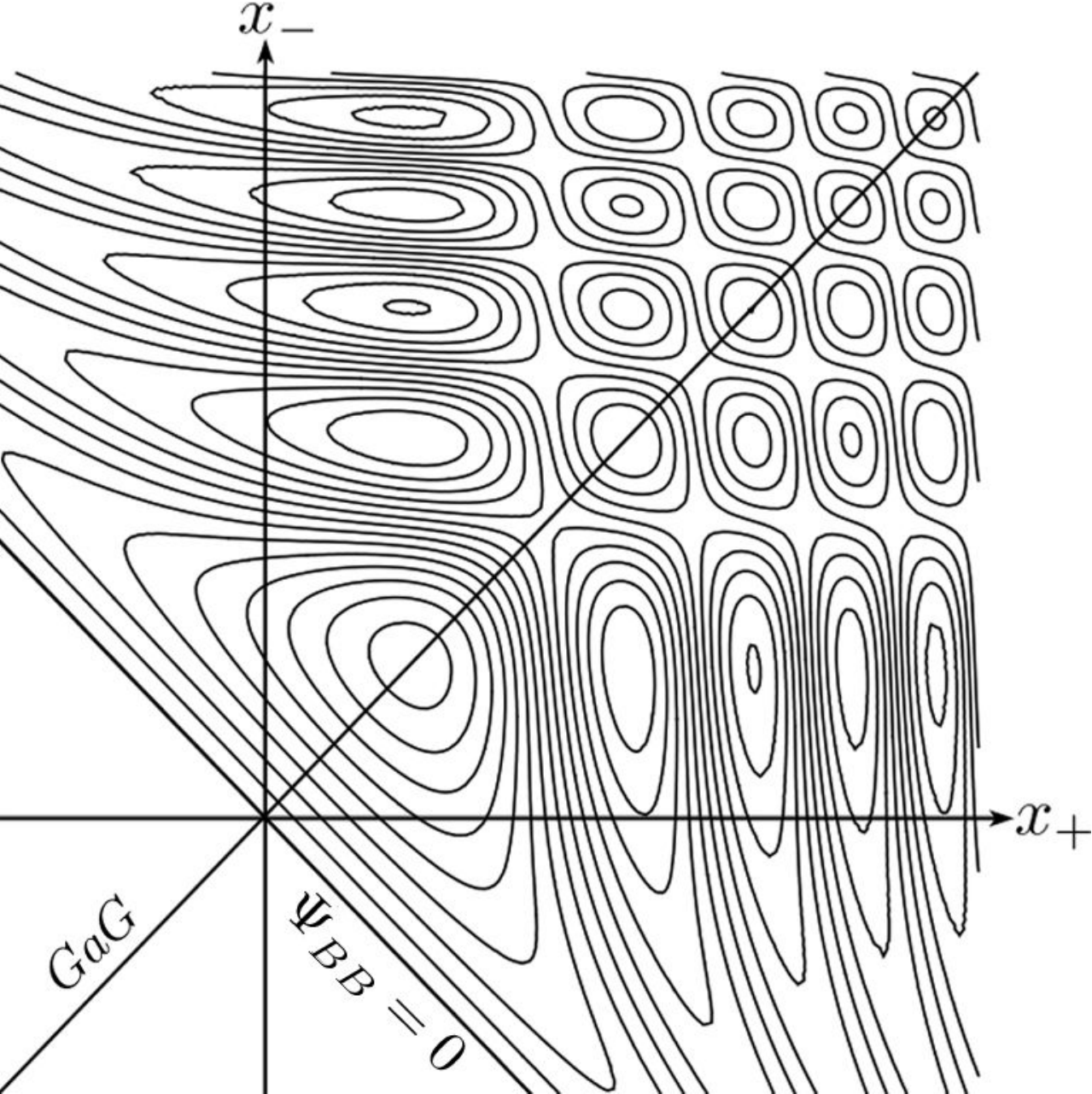}
	\caption{$|\Psi_S|^2$ contour plot (only valid for $k=0$,
	no $k=\pm 1$ extensions):
	$\Psi_S$ is GaG-even/BB-odd, so the $a^2<0$ sector
	can be dismissed.
	Notice the grid structure at the 1st quarter.}
	\label{ContourBB}
\end{figure}

\smallskip
\noindent \emph{(ii) Strong DeWitt initial condition}
$\psi_S (x,-x)=0$:
Here we further require that, as the cosmological scale factor
approaches zero, $\psi\rightarrow 0$ for any finite value of
$a\phi$ (infinite $\phi$ included).
Following the above prescription, one constructs the
GaG-even BB-odd entangled combination
\begin{equation}
	\Psi_S (x_{+},x_{-})= \frac{1}{\sqrt{2}}
	\left(R(x_+)L(-x_-)-R(-x_-)L(x_+)\right)
\end{equation}
but immediately realizes that this is a solution of the
WDW equation only provided $k=0$
(for which $U(-x)= U(-x)$).
It is remarkable how the strong DeWitt initial condition actually
singles out the flat space configuration.
Contrary to the previous case, it is now the $a^2<0$ sector which
can be dismissed.
The classical analog here is the $\{a^2(t)\sim
\cosh \omega t,~a(t)\phi(t)\sim \sinh\omega t\}$-type solution.
The $|\Psi_S|^2$ contour plot is depicted in Fig.(\ref{ContourBB}).

The first mathematical feature observed is the
two-dimensional grid formed by the local maxima of $|\Psi_{W,S}|^2$.
In what follows we analyze the grid structure of $\Psi_S$.
We start with the simplest special case $E=0$,
and hence $s(0)=0$ by virtue of $k=0$, for which the
large-$x_{\pm}$ behavior in the $x_{\pm}>0$ region is
proportional to
\begin{equation}
	\Psi_S\propto
	\sin \left[\frac{\sqrt{3\Lambda}}{\hbar}(x^2_++x^2_-)\right]
	+\frac{1}{\sqrt{2}}
	\cos \left[\frac{\sqrt{3\Lambda}}{\hbar}(x^2_+-x^2_-)\right]
\end{equation}
subject to an overall $(x_+ x_-)^{-1/2}$ suppression factor.
The density of zeroes grows  $\sim\sqrt{N}$, thus rapidly
approaching the classical regime.
The local extrema are located at
\begin{equation}
	x^2_{\pm}\approx\frac{h \left(n_{\pm}
	+\frac{1}{4}\right)}{2\sqrt{3\Lambda}}~,
\end{equation}
where $n_{\pm}$ are non-negative integers.
In turn, the locally most probable values for the
scale factor and the inverse Newton constant are well
approximated by
\begin{equation}
	\left[\begin{array}{c}a^2 \\a\phi\end{array}\right]
	\approx  \sqrt{\frac{h}{2\sqrt{3\Lambda}}}
	\left( \sqrt{n_++\frac{1}{4}} \pm \sqrt{n_-+\frac{1}{4}}\right)~.
\end{equation}
Numerically, the highest peak is the closest one to the origin,
located at $a_0=0.874\sqrt{\hbar}/\Lambda^{-1/4}$ along
the GaG axis.
In the other $x_+ x_-<0$ region, owing to the different asymptotic
behavior $\Psi_S\propto
\sin \left[\frac{\sqrt{3\Lambda}}{\hbar}(x^2_+-x^2_-)\right]$,
the grid turns one-dimensional, exhibiting an approximated
$a^3\phi$-periodicity.
Once the energy parameter $E$ is switched on, and hence
$s(E)=\frac{E}{\hbar}\sqrt{\frac{3}{\Lambda}}$, only mild
changes are experienced.
The major change has to do with a logarithmic modification of the
approximate periodicity $\Delta x^2$, that is
\begin{equation}
	\Delta \left(x^2+\frac{s(E)}{4\sqrt{3\Lambda}}
	\log \frac{2 (3\Lambda)^{\frac{1}{4}}x}{\sqrt{\hbar}}\right)=
	\frac{h}{\sqrt{3\Lambda}}~.
\end{equation}
Finally we note that the most general WdW wave function, subject to the
strong/weak DeWitt initial conditions, is of the form
$\int \Psi_{S,W}(E) f(E)~dE$ involving some weight function $f(E)$.

\medskip
\noindent\textbf{Epilog}

Recalling the underlying upside down effective potential $V_{eff}(\phi)
=-\frac{2}{3}\Lambda \phi^2$, the quantum mechanical background
associated with the BB-odd GaG-even WDW wave function $\Psi_S$
constitutes a perfect setup for a spontaneous GaG symmetry breaking
mechanism following a quantum creation at the top of the hill.
Appreciating the fact that a GaG-odd $V(\phi)$ gets generically
translated into a GaG-even $V_{eff}(\phi)$, one may envision a
cosmological Klein-Gordon evolution (entropy oriented arrow of
time?) governed by a double well $V_{eff}(\phi)$, with general
relativity formulated around the dilaton vacuum expectation value
$\langle\phi\rangle=(8\pi G)^{-1}$.
This holds for any $E$ in the spectrum, with potentially
far reaching consequences for the multiverse hypothesis.

\acknowledgments
{We cordially thank BGU president Prof. Rivka Carmi for her kind
support, and our colleagues Dr. Shimon Rubin and Ben Yellin for
many stimulating discussions.}

\end{document}